\documentclass[aps,preprint,amsmath,amssymb,amsfonts,showpacs]{revtex4}
\usepackage{epsfig}
\usepackage{graphicx}
\usepackage{subfigure}
\usepackage{dcolumn}
\usepackage{bm}

\newcommand{\del}{\partial}

\newcommand{\ddsimple}[2]{\frac{\del #1}{\del #2}} 

\newcommand{\bbR}{\mathbb{R}}
\newcommand{\bbC}{\mathbb{C}}
\newcommand{\bbN}{\mathbb{N}}
\newcommand{\bbP}{\mathbb{P}}
\newcommand{\bbT}{\mathbb{T}}
\newcommand{\bbZ}{\mathbb{Z}}
\newcommand{\scri}{\mathcal{I}}

\begin{document}

\title{Twistors and antipodes in de Sitter space}

\author{Yasha Neiman}
\email{yashula@gmail.com}
\affiliation{Institute for Gravitation \& the Cosmos and Physics Department, Penn State, University Park, PA 16802, USA}

\date{\today}

\begin{abstract}
We develop the basics of twistor theory in de Sitter space, up to the Penrose transform for free massless fields. We treat de Sitter space as fundamental, as one does for Minkowski space in conventional introductions to twistor theory. This involves viewing twistors as spinors of the de Sitter group $SO(4,1)$. When attached to a spacetime point, such a twistor can be reinterpreted as a local $SO(3,1)$ Dirac spinor. Our approach highlights the antipodal map in de Sitter space, which gives rise to doublings in the standard relations between twistors and spacetime. In particular, one can generate a field with both handedness signs from a single twistor function. Such fields naturally live on antipodally-identified de Sitter space $dS_4/\bbZ_2$, which has been put forward as the ideal laboratory for quantum gravity with positive cosmological constant.
\end{abstract}


\maketitle
\newpage

\section{Introduction} \label{sec:intro}

Observation suggests that our universe has a positive cosmological constant. This makes de Sitter space $dS_4$ the most physically relevant of the maximally symmetric spacetimes. In some ways, the theoretical understanding of this spacetime is also the least developed. Our ignorance is especially poignant when considering quantum gravity. Since the resources available to an observer in de Sitter space are always limited by the cosmological horizon, the fate of sharp observables becomes unclear. Indeed, one is no longer sure about the nature of the Hilbert space of states, e.g. whether or not it is observer-dependent. For discussions, see e.g. \cite{Witten:2001kn,Parikh:2002py,ArkaniHamed:2007ky}. The conceptual challenges for quantum gravity in de Sitter space closely mirror the more general ones concerning horizon thermodynamics and quantum gravity in finite spatial regions. Thus, in addition to its relevance to real-world cosmology, de Sitter space may serve as the simplest theoretical laboratory for exploring these issues.   

With the above motivation, it is of interest to adapt to de Sitter space every theoretical tool that was developed for the Minkowski or anti-de Sitter (AdS) settings. In the present paper, we aim to do this for the basics of twistor theory. Twistor theory \cite{Penrose:1986ca,Ward:1990vs} is an approach to geometry and physics that seeks to shift the focus from spacetime to twistor space - the spin-1/2 representation space of the spacetime symmetry algebra. Geometrically, a twistor is a totally-null plane in the complexified spacetime, while a spacetime point is a Riemann sphere in twistor space. The shift to twistor space lends greater importance to holomorphic structures, which ideally take over the role of spacetime field equations. In recent years, twistor theory was involved in significant advances in S-matrix calculations for $\mathcal{N}=4$ super-Yang-Mills \cite{Adamo:2011pv,Arkani-Hamed:2013jha} and for $\mathcal{N}=8$ supergravity \cite{Skinner:2013xp}. These advances suggest twistors as the optimal description for on-shell massless particles in Minkowski space. One can hope, then, that an improved understanding of twistors in de Sitter space may point towards the correct de Sitter substitute for the concept of particles.  

Certainly, de Sitter space is not new to the twistor literature. First, much of twistor theory is conformally invariant. This means that elementary twistor language, initially developed for Minkowski space, can be carried over to de Sitter through a conformal transformation. For a recent application in the context of modern S-matrix methods, see \cite{Adamo:2013tja}. Alternatively, one may view de Sitter space as a particular case of more general curved spacetimes, and then use more advanced methods such as a local twistor bundle. One goal of the present paper is to bridge the cosmetic gap left by these approaches. We aim to describe twistor theory in a way that uses the special structure of de Sitter space, and does so \emph{on its own terms}, without having to go through Minkowski space. Thus, the paper can be read as an unorthodox introduction to twistors, with de Sitter space instead of Minkowski as the starting point.
 
Cosmetics aside, there is an upshot to shifting the focus away from Minkowski space. Indeed, Minkowski space is not quite conformally equivalent to de Sitter, but to a patch that only covers half of $dS_4$. By focusing on such a patch, one loses conceptually important features of de Sitter space, such as observer-dependent cosmological horizons. A related feature is the antipodal map $x\rightarrow -x$, which always takes one out of the conformally flat patch and across the cosmological horizon. One may choose to topologically identify antipodal points, which yields the so-called ``elliptical'' de Sitter space $dS_4/\bbZ_2$. This spacetime should not be confused with the (geodesically incomplete) half of $dS_4$ that is conformal to Minkowski space. The peculiar properties of the quotient space $dS_4/\bbZ_2$ are reviewed in \cite{Parikh:2002py}. It is argued there and elaborated in \cite{Parikh:2004ux,Parikh:2004wh} that $dS_4/\bbZ_2$ is in fact a more promising setting for quantum gravity with positive cosmological constant than $dS_4$ itself.

Thus, another aim of this paper is to study the antipodal map and the space $dS_4/\bbZ_2$ in twistor language. We find that the antipodal map induces certain doublings in the standard relations between twistor space and spacetime. In particular, a twistor is now no longer a left-handed totally-null plane in complexified $dS_4$, but a pair of left-handed and right-handed planes. Conversely, a point in $dS_4$ is not one Riemann sphere in twistor space, but two. This leads to a version of the Penrose transform that produces both self-dual and anti-self-dual massless free fields from the same type of twistor function. In particular, one can use it to produce fields on $dS_4/\bbZ_2$, where there is \emph{no} global notion of self-duality.

In our construction, the infinity twistor plays a peculiar role. Normally, this is the structure in twistor theory that breaks the conformal group down to the group of isometries. The antipodal map in $dS_4$ is invariant under the isometry group $SO(4,1)$, but it is not conformally invariant - it isn't part of the causal structure. Accordingly, we will see that in twistor language, the antipodal map involves the infinity twistor. On the other hand, in the quotient space $dS_4/\bbZ_2$, the antipodal map becomes incorporated into the \emph{global} causal structure. This reflects the fact that for topological reasons, the conformal group on $dS_4/\bbZ_2$ is no bigger than the isometry group. Since the relation between twistors and spacetime is non-local, such global features are significant. Thus, in $dS_4/\bbZ_2$, the infinity twistor becomes a legitimate part of the conformal structure.     

The rest of the paper is organized as follows. In section \ref{sec:dS}, we introduce de Sitter space in terms of its embedding in the 4+1d flat space $\bbR^{4,1}$. In section \ref{sec:clifford}, we introduce twistors as the spinors of $SO(4,1)$. In section \ref{sec:bilinears}, we outline the geometry of twistors in de Sitter space. Conversely, in section \ref{sec:projectors}, we outline the geometry of spacetime points in twistor space. We also describe there how twistors can be ``evaluated'' at a spacetime point to yield local Dirac spinors. In section \ref{sec:scalar}, we present the twistor transform for a conformally coupled scalar field. In section \ref{sec:spin}, we present the transforms for free massless fields with spin. Section \ref{sec:discuss} is devoted to discussion and outlook. 

We use indices $(\mu,\nu,\dots)=0,1,2,3,4$ to denote vectors in 4+1d flat space. Their projections onto the de Sitter hyperboloid will be identified with intrinsic vectors in $dS_4$. We use 4d indices $(a,b,\dots)$ for twistors, as well as for local Dirac spinors in de Sitter space. In the latter case, the index is a direct sum of a left-handed Weyl spinor index $(\alpha,\beta,\dots)$ and a right-handed one $(\dot\alpha,\dot\beta,\dots)$.

\section{Variants of de Sitter space: global, compactified, elliptical and complex} \label{sec:dS}

De Sitter space $dS_4$ can be modeled as the hyperboloid $x_\mu x^\mu = 1$ of unit spacelike radius in the 4+1d Minkowski space $\bbR^{4,1}$. Alternatively, it is the set of spacelike \emph{directions} in $\bbR^{4,1}$. The future and past conformal infinities $\scri^\pm$ can be viewed as the 3-spheres of future-pointing and past-pointing null directions in $\bbR^{4,1}$, respectively. One may choose to identify pairs of points related through the antipodal map $x^\mu \leftrightarrow -x^\mu$. This takes us to ``elliptical'' de Sitter space $dS_4/\bbZ_2$. The conformal infinities $\scri^\pm$ are then identified into a single 3-sphere $\scri$. Alternatively, one can identify the $\scri^\pm$ without identifying antipodal points in the bulk. This results in compactified de Sitter space. 

Similarly, complex de Sitter space $dS_{4,\bbC}$ is the set of points with $x_\mu x^\mu = 1$ in the flat complex space $\bbC^5$. Again, we have an optional identification of antipodal points, which takes us to $dS_{4,\bbC}/\bbZ_2$. If we view complex de Sitter space as the set of non-null complex \emph{directions} in $\bbC^5$ (there is no longer a distinction between spacelike and timelike), then the identification of antipodal points becomes mandatory: the directions of $x^\mu$ and $-x^\mu$ are continuously related through phase rotations $x\rightarrow e^{i\theta}x$. For the same reason, whether or not we identify antipodal points in the bulk, we \emph{must} identify the two infinities $\scri^\pm$ when complexifying. The resulting complexified infinity $\scri_\bbC$ is the set of complex null directions in $\bbC^5$. 

Like Minkowski space, $dS_4$ is separately orientable in space and time. Together, these define the spacetime orientation, captured by the Levi-Civita symbol $\epsilon^{\mu\nu\rho\sigma}$. The latter is related to the Levi-Civita symbol in $\bbR^{4,1}$ as:
\begin{align}
 \epsilon^{\mu\nu\rho\sigma} = \epsilon^{\mu\nu\rho\sigma\lambda}x_\lambda \ . \label{eq:epsilons}
\end{align}
Since the antipodal map $x^\mu \rightarrow -x^\mu$ reverses the time direction, elliptical de Sitter space $dS_4/\bbZ_2$ is not time-orientable. Moreover, since $\bbR^{4,1}$ has an even spatial dimension and an odd spacetime dimension, \eqref{eq:epsilons} implies that $dS_4/\bbZ_2$ inherits a spatial orientation from $\bbR^{4,1}$, but not a spacetime one. This will be important below, when we consider the handedness of free massless fields. Similarly, complex de Sitter space $dS_{4,\bbC}$ has a spacetime orientation, while $dS_{4,\bbC}/\bbZ_2$ does not. As always with complex spacetimes, neither of the two has a separate notion of spatial or time orientation.

\section{Twistors and the $SO(4,1)$ Clifford algebra} \label{sec:clifford}

Twistors are often defined as the Weyl spinors of the spacetime conformal group $SO(4,2)$. In Minkowski space, the conformal group can be reduced to the Poincare group by introducing the infinity twistor $I_{ab}$. In de Sitter space, a different (non-degenerate) choice of $I_{ab}$ reduces us instead to the de Sitter isometry group $SO(4,1)$. We view this as the group of rotations in the embedding flat space $\bbR^{4,1}$. We then introduce twistor space $\bbT$ as the space of Dirac spinors of $SO(4,1)$.

In the embedding flat space $\bbR^{4,1}$, we have the 4+1d Clifford algebra, generated by the gamma matrices $(\gamma^\mu)^a{}_b$. These can be represented in $2\times 2$ block notation as:
\begin{align}
 (\gamma^0)^a{}_b = \begin{pmatrix} 0 & 1 \\ 1 & 0 \end{pmatrix} \ ; \quad 
 (\gamma^k)^a{}_b = \begin{pmatrix} \tau^k & 0 \\ 0 & -\tau^k \end{pmatrix} \ ; \quad 
 (\gamma^4)^a{}_b = \begin{pmatrix} 0 & -1 \\ 1 & 0 \end{pmatrix} \ , \label{eq:gamma}
\end{align}
where the $2\times 2$ matrices $\tau_k \equiv -i\sigma_k$ for $k=1,2,3$ are imaginary multiples of the Pauli matrices:
\begin{align}
 \tau_1 = \begin{pmatrix} 0 & -i \\ -i & 0 \end{pmatrix} \ ; \quad
 \tau_2 = \begin{pmatrix} 0 & -1 \\ 1 & 0 \end{pmatrix} \ ; \quad
 \tau_3 = \begin{pmatrix} -i & 0 \\ 0 & i \end{pmatrix} \ . \label{eq:tau}
\end{align}
The $\tau_k$ satisfy the quaternionic algebra $\tau_i\tau_j = -\delta_{ij} + \epsilon_{ijk}\tau^k$. The gamma matrices \eqref{eq:gamma} satisfy the Clifford algebra $\gamma^{(\mu}\gamma^{\nu)} = -\eta^{\mu\nu}$, where $\eta^{\mu\nu}$ is the inverse of the flat 4+1d metric $\eta_{\mu\nu}$ with mostly-plus signature. Note that in treatments of $SO(3,1)$, our $\gamma^4$ is usually denoted as $\gamma^5$. The antisymmetric products $\gamma^{[\mu}\gamma^{\nu]} \equiv \gamma^{\mu\nu}$ are given by:
\begin{align}
 \begin{split}
   &(\gamma^{04})^a{}_b = \begin{pmatrix} 1 & 0 \\ 0 & -1 \end{pmatrix} \ ; \quad
   (\gamma^{ij})^a{}_b = \epsilon^{ijk} \begin{pmatrix} \tau_k & 0 \\ 0 & \tau_k \end{pmatrix} \ ; \\  
   &(\gamma^{0k})^a{}_b = \begin{pmatrix} 0 & -\tau^k \\ \tau^k & 0 \end{pmatrix} \ ; \quad 
   (\gamma^{4k})^a{}_b = \begin{pmatrix} 0 & \tau^k \\ \tau^k & 0 \end{pmatrix} \ . \\ 
 \end{split} \label{eq:gamma_mn}
\end{align}
The $SO(4,1)$ group is generated by the matrices $\gamma^{\mu\nu}/2$. As an aside, the absence of explicit $i$ factors in \eqref{eq:gamma} and \eqref{eq:gamma_mn} allows us to interpret $(\gamma^\mu,\gamma^{\mu\nu})$ as $2\times 2$ matrices over the quaternions. This reflects the fact that the double cover of $SO(4,1)$ is the quaternionic group $Sp(1,1)$. The product of three gamma matrices is another gamma matrix. In particular, we have $\gamma^{[\mu}\gamma^\nu\gamma^\rho\gamma^\sigma\gamma^{\lambda]} = \epsilon^{\mu\nu\rho\sigma\lambda}$, where $\epsilon^{\mu\nu\rho\sigma\lambda}$ is the Levi-Civita symbol with $\epsilon^{01234} = 1$. Both $\gamma^\mu$ and $\gamma^{\mu\nu}$ are traceless. Together with the unit matrix, they span the $4\times 4$ matrix space. Useful trace identities include:
\begin{align}
 \delta^a_a = 4 \ ; \quad (\gamma^\mu)^a{}_a = 0 \ ; \quad (\gamma^\mu)^a{}_b(\gamma_\nu)^b{}_a = -4\delta^\mu_\nu \ ;\quad 
 (\gamma^{\mu\nu})^a{}_b(\gamma_{\rho\sigma})^b{}_a = -8\delta^{[\mu}_{[\rho}\delta^{\nu]}_{\sigma]} \ .
\end{align}
With these, one can decompose any twistor matrix $M^a{}_b$ into 4+1d scalar, vector and bivector pieces:
\begin{align}
 \begin{split}
   &M^a{}_b = m\delta^a_b + m^\mu \gamma_\mu{}^a{}_b + m^{\mu\nu}\gamma_{\mu\nu}{}^a{}_b \ ; \\
   &m = \frac{1}{4}M^a{}_a \ ; \quad m^\mu = -\frac{1}{4}M^a{}_b \gamma^{\mu b}{}_a \ ; \quad m^{\mu\nu} = -\frac{1}{8}M^a{}_b \gamma^{\mu\nu b}{}_a \ .
 \end{split} \label{eq:decompose}
\end{align}
The $SO(4,1)$ group leaves invariant the antisymmetric ``infinity twistor'':
\begin{align}
 I_{ab} = -I_{ba} = \begin{pmatrix} 0 & \tau_2 \\ \tau_2 & 0 \end{pmatrix} = I^{ab} = -I^{ba} \ ; \quad I_{ac}I^{bc} = \delta_a^b \ .
\end{align}
We can use $I_{ab}$ and $I^{ab}$ to raise and lower twistor indices as:
\begin{align}
 Z_a = I_{ab}Z^b \quad ; \quad Z^a = Z_b I^{ba} \ . \label{eq:raise_lower}
\end{align}
In particular, we see that $I_{ab}$ and $I^{ab}$ are indeed lowered/raised-index versions of each other. We also have:
\begin{align}
 I^a{}_b = -I_b{}^a = \delta_b^a \ .
\end{align}
Lowering indices on \eqref{eq:gamma} and \eqref{eq:gamma_mn}, we find that the $\gamma^\mu_{ab}$ are antisymmetric (and traceless with respect to $I^{ab}$), while the $\gamma^{\mu\nu}_{ab}$ are symmetric:
\begin{align}
 \gamma^\mu_{ab} = -\gamma^\mu_{ba} \ ; \quad I^{ab}\gamma^\mu_{ab} = 0 \ ; \quad \gamma^{\mu\nu}_{ab} = \gamma^{\mu\nu}_{ba} \ .
\end{align}
The six matrices $(I_{ab},\gamma^\mu_{ab})$ span the antisymmetric $4\times 4$ matrix space, while the ten matrices $\gamma^{\mu\nu}_{ab}$ span the symmetric one. 
Other useful identities include:
\begin{align}
 &\epsilon^{abcd} = -3I^{[ab}I^{cd]} \ ; \quad \epsilon^{abcd}I_{cd} = -2I^{ab} \ ; \quad \epsilon^{abcd}\gamma^\mu_{cd} = 2\gamma^{\mu ab} \ ; \\
 &\gamma_\mu^{ab}\gamma^\mu_{cd} = I^{ab}I_{cd} - 4\delta^{[a}_{[c} \delta^{b]}_{d]} \ . \label{eq:gamma_gamma}
\end{align}
If we restrict to \emph{real} $SO(4,1)$ rotations, an additional invariant structure appears - a Hermitian metric with signature $(2,2)$:
\begin{align}
 \bar Z_a = \begin{pmatrix} 0 & -1 \\ -1 & 0 \end{pmatrix} \overline{Z^b} \quad ; \quad
 \bar Z^a = \begin{pmatrix} 0 & 1 \\ 1 & 0 \end{pmatrix} \overline{Z_b} \ , \label{eq:Hermitian}
\end{align}
where the signs are chosen so that \eqref{eq:raise_lower} and \eqref{eq:Hermitian} commute. The Hermitian metric \eqref{eq:Hermitian} identifies the de Sitter group $SO(4,1)\approx Sp(1,1)$ as a subgroup of the conformal group $SO(4,2)\approx SU(2,2)$. Together, \eqref{eq:raise_lower} and \eqref{eq:Hermitian} can be combined into a complex-conjugation operation:
\begin{align}
 \bar Z^a = \begin{pmatrix} \tau_2 & 0 \\ 0 & \tau_2 \end{pmatrix} \overline{Z^b} \quad ; \quad
 \bar Z_a = \begin{pmatrix} \tau_2 & 0 \\ 0 & \tau_2 \end{pmatrix} \overline{Z_b} \ . \label{eq:conjugate}
\end{align}
This complex conjugation is anti-idempotent and commutes with scalar products: 
\begin{align}
 \Bar{\Bar Z}^a = -Z^a \quad ; \quad \bar W_a \bar Z^a = \overline{W_a Z^a} \ .
\end{align}
Due to the anti-idempotence, there are no real twistors. Moreover, $Z^a$ and $\bar Z^a$ are always linearly independent. On the other hand, the special twistor matrices introduced above are all real under \eqref{eq:conjugate}:
\begin{align}
 \bar I_{ab} = I_{ab} \ ; \quad \bar\gamma^\mu_{ab} = \gamma^\mu_{ab} \ ; \quad \bar\gamma^{\mu\nu}_{ab} = \gamma^{\mu\nu}_{ab} \ . 
\end{align}
We see that in the de Sitter context, there is no distinction between twistor space and its dual: twistor indices can be raised, lowered and conjugated freely.

\section{The geometry of twistor bilinears} \label{sec:bilinears}

\subsection{Complex and real bilinears from a twistor $Z^a$}

In complex Minkowski space, a projective twistor $Z^a$ (i.e. a twistor defined up to rescalings $Z^a\rightarrow \lambda Z^a$) corresponds to a totally null 2-plane of a certain handedness, known as an $\alpha$-plane. If the twistor is null, i.e. $\bar Z_a Z^a = 0$, then its $\alpha$-plane intersects real spacetime at a lightray. Let us find the analogues of these results in de Sitter space.

Given a twistor $Z^a$, we can construct the bilinear $Z^a Z_b$. This can be decomposed into 4+1d scalar, vector and bivector pieces, according to \eqref{eq:decompose}. Since $Z^a Z^b$ is symmetric, only the bivector piece will be non-vanishing. It's easy to check that the bivector corresponding to $Z^a Z_b$ is simple and totally null. See table \ref{tab:bilinears} for examples. The bivector's projective version, taking into account the freedom to rescale $Z^a$, is a totally null 2-plane through the origin in $\bbC^5$. This translates into a null geodesic at complexified de Sitter infinity $\scri_\bbC$. Thus, projective twistor space $\bbP\bbT$ is the space of null geodesics on $\scri_\bbC$. For a null twistor $Z^a$, the null geodesic at $\scri_\bbC$ intersects the real 3-sphere $\scri$ at a point. This corresponds to two antipodally related points on $\scri^+$ and $\scri^-$.
\begin{table}
 \centering
 \begin{tabular}{|c|c|c|c|} 
  \hline
  $Z^a$ & Region & $Z^a Z_b$ & $Z^a\bar Z_b$ \\
  \hline
  $(1,0,0,0)$ & $\bbN^{\hphantom{+}}$ & $\frac{i}{4}(\gamma_0 + \gamma_4)(\gamma_1 + i\gamma_2)$ & $\frac{1}{4}(\gamma_0 + \gamma_4)(1 - i\gamma_3)$ \\
  \hline
  $(1,0,-1,0)$ & $\bbT^+$ & $-\frac{1}{2}(\gamma_1 + i\gamma_2)(\gamma_3 + i\gamma_4)$ & $\frac{1}{2}(+1 + \gamma_0 + i(\gamma_{12} + \gamma_{34}))$ \\
  \hline
  $(1,0,1,0)$ & $\bbT^-$ & $\hphantom{-}\frac{1}{2}(\gamma_1 + i\gamma_2)(\gamma_3 - i\gamma_4)$ & $\frac{1}{2}(-1 + \gamma_0 - i(\gamma_{12} - \gamma_{34}))$ \\
  \hline
 \end{tabular}
 \caption{Examples of twistors $Z^a$ from different regions in twistor space. For each twistor, we present the 4+1d Clifford algebra elements corresponding to the bilinears $Z^a Z_b$ and $Z^a\bar Z_b$.}
 \label{tab:bilinears}
\end{table}

Alternatively, one can dualize the 2-plane in $\bbC^5$ corresponding to $Z^a Z_b$, obtaining a 3-plane with a rank-1 intrinsic metric. This 3-plane intersects complexified de Sitter space $dS_{4,\bbC}$ at a pair of totally null 2-planes, related through the antipodal map. The pair of 2-planes intersect each other and $\scri_\bbC$ at the null geodesic described above. Since we've seen that the antipodal map reverses spacetime orientation, the two 2-planes have opposite handedness. Thus, they constitute an $\alpha$-plane and $\beta$-plane pair. When the twistor $Z^a$ is null, the 2-planes intersect the real spacetime $dS_4$. In that case, the intersection is a pair of antipodally related lightrays.

If complex conjugation is allowed, we can consider also the bilinear $Z^a\bar Z_b$. We again apply the decomposition \eqref{eq:decompose} into 4+1d scalar, vector and bivector pieces. This time, all the pieces are non-vanishing in general. Under the complex conjugation $Z^a\rightarrow \bar Z^a$, the scalar and vector pieces of $Z^a\bar Z_b$ are invariant, while the bivector piece changes sign.

The scalar piece of $Z^a\bar Z_b$ is the real scalar $\bar Z_a Z^a/4$. Projectively, one can only say whether it's positive, negative or zero. The three possibilities divide twistor space into regions $\bbT^+$, $\bbT^-$ and $\bbN$, as usual. See table \ref{tab:bilinears} for examples. The twistor complex conjugation \eqref{eq:conjugate} maps each region onto itself.

The vector piece of $Z^a \bar Z_b$ is real and future-pointing. It is null for $Z^a\in \bbN$ and timelike for $Z^a\in \bbT^\pm$. Projectively, it defines a causal line through the origin in $\bbR^{4,1}$. This is the line orthogonal to the bivector $\gamma^{\mu\nu}_{ab}Z^a Z^b$ and to its complex conjugate $\gamma^{\mu\nu}_{ab}\bar Z^a \bar Z^b$. For $Z^a\in \bbN$, the (null) vector piece of $Z^a \bar Z_b$ defines a point on $\scri^+$, along with its antipode on $\scri^-$. These are the same as the points defined by $Z^a Z_b$.
 
The bivector piece of $Z^a \bar Z_b$ is imaginary. Along with the bivectors corresponding to $Z^a Z_b$ and $\bar Z^a \bar Z_b$, it lives in the 4-plane orthogonal to the vector piece of $Z^a \bar Z_b$. For $Z^a\in \bbT^+$, this 4-plane is spacelike, and the three bivectors are all self-dual with respect to its future-pointing normal. Similarly, for $Z^a\in \bbT^-$, the bivectors are all anti-self-dual with respect to the future-pointing normal. For $Z^a\in \bbN$, the bivector piece of $Z^a \bar Z_b$ is simple, with signature $(0,+)$. Projectively, it defines a $(0,+)$-signature 2-plane through the origin in $\bbR^{4,1}$. This 2-plane intersects $dS_4$ at a pair of lightrays, related through the antipodal map. These are the same as the two lightrays arising from $Z^a Z_b$. The lightrays begin and end at the points on $\scri^\pm$ defined by $Z^a Z_b$ or by the vector piece of $Z^a \bar Z_b$. 

\subsection{Summary of structures in de Sitter space associated with a twistor}

\noindent
Projective twistor space $\bbP\bbT$ is:
\begin{itemize}
 \item The space of null geodesics at complexified infinity $\scri_\bbC$.
 \item The space of totally null 2-surfaces in $dS_{4,\bbC}/\bbZ_2$ (with no definite handedness).
 \item The space of antipodally related \emph{pairs} of totally null 2-surfaces in $dS_{4,\bbC}$, with opposite handedness signs.
\end{itemize}
Projective null twistor space $\bbP\bbN$ is a \emph{double cover} of: 
\begin{itemize}
 \item The space of lightrays in $dS_4/\bbZ_2$.
 \item The space of antipodally related \emph{pairs} of lightrays in $dS_4$.
\end{itemize}
The double cover is due to the fact that a null twistor $Z^a$ and its (linearly independent) complex conjugate $\bar Z^a$ correspond to the same pair of lightrays in $dS_4$.

\section{Spacetime points, Riemann spheres and chiral projectors} \label{sec:projectors}

\subsection{The projectors $P_\pm(x)$}

In section \ref{sec:bilinears}, we identified the structures in de Sitter space associated with a (projective) twistor. Let us now address the converse question - what structures in twistor space are associated with a de Sitter point? In complex Minkowski space, a spacetime point maps to a Riemann sphere $\bbC\bbP^1\subset \bbP\bbT$, with real points mapped to Riemann spheres in $\bbP\bbN$. Let us see how things change in the de Sitter case.

Consider a point in $dS_{4,\bbC}$, parametrized by a 4+1d complex vector $x^\mu$ with $x_\mu x^\mu = 1$. Define the following pair of twistor matrices:
\begin{align}
 P_\pm{}^a{}_b(x) = \frac{1}{2}(\delta^a_b \pm ix^\mu\gamma_\mu{}^a{}_b) \ , \label{eq:projectors} 
\end{align}
with raised-index versions:
\begin{align}
 P_\pm^{ab}(x) = \frac{1}{2}(I^{ab} \pm ix^\mu\gamma_\mu^{ab}) \ . \label{eq:P_upper}
\end{align}
It's easy to check that the matrices $P_\pm^{ab}(x)$ are antisymmetric and simple, i.e. rank-2. Thus, they define a pair of $\bbC^2$ subspaces in $\bbT$, or Riemann spheres in $\bbP\bbT$. We denote both the subspaces and the Riemann spheres as $P_\pm(x)$. The matrices $P_\pm{}^a{}_b$ are projectors onto the respective $\bbC^2$ subspaces. They satisfy:
\begin{align}
 P_-{}^a{}_c P_-{}^c{}_b = P_-{}^a{}_b \ ; \quad P_+{}^a{}_c P_+{}^c{}_b = P_+{}^a{}_b \ ; \quad  
 P_-{}^a{}_c P_+{}^c{}_b = P_+{}^a{}_c P_-{}^c{}_b = 0 \ ,
\end{align}
where the last equation implies $W_a Y^a = 0$ for all $W^a\in P_-(x)$ and $Y^a\in P_+(x)$. The $P_-(x)$ and $P_+(x)$ projectors sum to unity. The gamma matrices can be decomposed in terms of $P_\pm(x)$ as follows:
\begin{align}
 x^\mu\gamma_\mu^{ab} &= i(P_-^{ab} - P_+^{ab}) \ ; \label{eq:gamma_5} \\ 
 (\delta_\mu^\nu - x_\mu x^\nu)\gamma_\nu^{ab} &= 2P_-{}^{[a}{}_c\, P_+{}^{b]}{}_d\, \gamma_\mu^{cd} \ . \label{eq:gamma_4}
\end{align}
Under the antipodal map $x^\mu\rightarrow -x^\mu$, the $P_\pm$ switch roles:
\begin{align}
 P_\pm^{ab}(-x) = P_\mp^{ab}(x) \ . \label{eq:P_antipodal}
\end{align}
Under the complex conjugation \eqref{eq:conjugate}, we have:
\begin{align}
 \bar P_\pm^{ab}(x) = P^{ab}_\mp(\bar x) \ . \label{eq:P_conjugate}
\end{align}
Thus, for real points $x^\mu$, the matrices $P_+^{ab}(x)$ and $P_-^{ab}(x)$ are complex conjugates. In this case, each of the subspaces $P_\pm(x)$ sits entirely in $\bbN$. Thus, $P_-(x)$ is a 2d subspace of null twistors $Z^a$, while $P_+(x)$ is the 2d subspace of their complex conjugates $\bar Z^a$. At the same time, it's not \emph{necessary} to use complex conjugation when discussing the $P_\pm(x)$: one can always use eq. \eqref{eq:projectors}, which is holomorphic in $x^\mu$. 

In the geometric language of section \ref{sec:bilinears}, the two Riemann spheres $P_\pm(x)$ correspond to the spheres of right-handed/left-handed totally null 2-planes in $dS_{4,\bbC}/\bbZ_2$ passing through the point $\pm x^\mu$. The distinction between left-handed and right-handed 2-planes can only be made locally in $dS_{4,\bbC}/\bbZ_2$, and there's no global way to decide which of the $P_\pm(x)$ corresponds to which handedness. For real points $x^\mu$, one can view \emph{either} $P_-(x)$ or $P_+(x)$ as the sphere of lightrays in $dS_4/\bbZ_2$ through $\pm x^\mu$: the null twistors in $P_-(x)$ map to the same lightrays as their complex conjugates in $P_+(x)$. 

Eq. \eqref{eq:P_upper} covers all the $\bbC^2$ subspaces of $\bbT$ except those on which the form $I_{ab}$ vanishes. The latter are given by simple antisymmetric matrices of the form $\ell^\mu\gamma_\mu^{ab}$, where $\ell^\mu$ is null. Since the $\bbC^2$ subspace does not depend on the scaling of $\ell^\mu$, these subspaces are in one-to-one correspondence with points at $\scri_\bbC$. 

\subsection{Summary of structures in twistor space associated with spacetime points}

\begin{itemize}
 \item Complex de Sitter space $dS_{4,\bbC}$ is the space of Riemann spheres in $\bbP\bbT$. The identification can be realized through either of the maps $P_\pm(x)$.
 \item $dS_{4,\bbC}/\bbZ_2$ is the space of unordered pairs $P_\pm$ of Riemann spheres in $\bbP\bbT$, whose bitwistors $P_\pm^{ab}$ can be scaled to satisfy $P_-^{ab} + P_+^{ab} = I^{ab}$. Alternatively, it is the space of unordered pairs of projectors $P_\pm{}^a{}_b$ that satisfy $P_-{}^a{}_b + P_+{}^a{}_b = \delta^a_b$ and $P_\pm{}^a{}_c I^{cb} = -P_\pm{}^b{}_c I^{ca}$. 
 \item The real spaces $dS_4$ and $dS_4/\bbZ_2$ are the spaces of Riemann spheres as above, on which the Hermitian metric \eqref{eq:Hermitian} vanishes.
 \item $\scri_\bbC$ is the space of Riemann spheres in $\bbP\bbT$ on which the form $I_{ab}$ vanishes.
 \item $\scri$ is the space of Riemann spheres on which both $I_{ab}$ and the Hermitian metric vanish.
\end{itemize}

\subsection{Dirac and Weyl spinors at a spacetime point} \label{sec:projectors:spinors}

The $\bbC^2$ subspaces $P_\pm(x)$ have an additional interpretation, which has no analogue in the twistor theory of Minkowski space. The projectors \eqref{eq:projectors} are just $x^\mu$-dependent versions of the familiar chiral projectors $(1\pm i\gamma_5)/2$ for $SO(3,1)$ Dirac spinors. Indeed, choosing a point $x^\mu$ breaks the $SO(4,1)$ de Sitter group down to the Lorentz group $SO(3,1)$ around the point. This induces a decomposition of twistor space (i.e. the Dirac representation of $SO(4,1)$) into the two Weyl representations of $SO(3,1)$. We can thus identify the $\bbC^2$ subspaces $P_\pm(x)$ with the left-handed and right-handed spin spaces at $x^\mu$! Again, if antipodal points are identified, then the left/right distinction only makes sense locally. 

We conclude that at a point $x^\mu$, a twistor index $a$ can be interpreted as a local $SO(3,1)$ Dirac index. This can then be viewed as a direct sum of a left-handed Weyl index $\alpha$ living in $P_-(x)$ and a right-handed index $\dot\alpha$ living in $P_+(x)$. The matrices $P_-^{ab}$ and $P_+^{ab}$ become the spinor metrics $\epsilon^{\alpha\beta}$ and $\epsilon^{\dot\alpha\dot\beta}$. The gamma matrices \eqref{eq:gamma_4} tangent to the de Sitter hyperboloid become the ordinary 3+1d gamma matrices, with components $\gamma_\mu^{\alpha\dot\alpha} = -\gamma_\mu^{\dot\alpha\alpha}$. The fifth, radial gamma matrix \eqref{eq:gamma_5} plays the role of $\gamma_5$, with components $i\epsilon^{\alpha\beta}$ and $-i\epsilon^{\dot\alpha\dot\beta}$. The canonical map between 3+1d tangent vectors $v^\mu$ and spinor matrices $v^{\alpha\dot\alpha}$ takes the form:
\begin{align}
 v^{\alpha\dot\alpha} = v^\mu\gamma_\mu^{\alpha\dot\alpha} \quad ; \quad 
 v^\mu = -\frac{1}{2}\,v^{\alpha\dot\alpha}\gamma^\mu_{\alpha\dot\alpha} \ , \label{eq:vectors_spinors}
\end{align}
where we used the identities:
\begin{align}
 \gamma^\mu_{\alpha\dot\alpha}\gamma_\nu^{\alpha\dot\alpha} = -2\delta^\mu_\nu \quad ; \quad
 \gamma_\mu^{\alpha\dot\alpha}\gamma^\mu_{\beta\dot\beta} = - 2\delta^\alpha_\beta \delta^{\dot\alpha}_{\dot\beta} \ ,
\end{align}
which follow from $\gamma^\mu_{ab}\gamma_\nu^{ab} = -4\delta^\mu_\nu$ and $\gamma_\mu^{ab}\gamma^\mu_{cd} = I^{ab}I_{cd} - 4\delta^{[a}_{[c} \delta^{b]}_{d]}$. Under the map \eqref{eq:vectors_spinors}, the de Sitter metric (i.e. the pullback of $\eta_{\mu\nu}$ onto the hyperboloid) is identified with $-2\epsilon_{\alpha\beta}\epsilon_{\dot\alpha\dot\beta}$.

The complex conjugation \eqref{eq:P_conjugate} interchanges the left-handed and right-handed Weyl spinors, as it should. Also, eq. \eqref{eq:P_antipodal} defines an isomorphism between left-handed spinors at $x$ and right-handed spinors at $-x$: the same twistor $Z^a$ can be viewed as either kind of Weyl spinor, depending on the spacetime point where it is ``evaluated''.   

The above construction allows us to view twistor-valued fields in de Sitter space as ordinary Dirac spinor fields. We will use this in the twistor transforms of section \ref{sec:spin}.

\section{Conformally coupled scalar field} \label{sec:scalar}
 
Before moving on to fields with spinor indices, let us work out the twistor transform for a free massless scalar field. More precisely, the field equation that arises naturally from the twistor transform is that of the \emph{conformally coupled} scalar:
\begin{align}
 \Box\varphi = \frac{1}{6}R\varphi = 2\varphi \ , \label{eq:scalar_eq}
\end{align} 
where we substituted the Ricci scalar $R = 12$ for de Sitter space with unit radius. The d'Alembertian in \eqref{eq:scalar_eq} is, of course, the covariant one for the curved space $dS_4$. However, one can substitute it with the flat d'Alembertian in $\bbR^{4,1}$, if we give $\varphi$ a trivial radial dependence $\varphi(x^\mu) = \varphi(\lambda x^\mu)$ for $\lambda$ in a neighborhood of $1$. It will be convenient to give this radial dependence also to the projectors \eqref{eq:projectors}, defining them for $x^\mu$ away from the hyperboloid $x_\mu x^\mu = 1$ as:
\begin{align}
 P_\pm{}^a{}_b(x) = \frac{1}{2}\left(\delta^a_b \pm \frac{ix^\mu}{\sqrt{x\cdot x}}\,\gamma_\mu{}^a{}_b \right) \ . \label{eq:projectors_full}
\end{align}
The ambiguity of the square root doesn't bother us, since we are only interested in a neighborhood of $x\cdot x = 1$.

Now, consider $\bar\del$-closed $(0,1)$-forms $f(Z)$ in twistor space, homogeneous of degree $-2$ and defined up to exact forms $f \rightarrow f + \bar\del h$. The space of such forms is spanned by the distributional ``elementary'' forms: 
\begin{align}
 f(Z) = \frac{1}{A_a Z^a}\, \bar\del\, \frac{1}{B_b Z^b} \ , \label{eq:f_elem}
\end{align}
where $A^a,B^a$ are a pair of constant twistors. Given two forms $f_\pm(Z)$ in this space, we define the twistor transform as:
\begin{align}
 \varphi(x) = \int_{P_-(x)} f_-(Z)\wedge Z_a dZ^a \,+\, \int_{P_+(x)} f_+(Z)\wedge Z_a dZ^a \ , \label{eq:scalar_transform}
\end{align}
where the integrals are over the Riemann spheres $P_\pm(x)$. The field $\varphi(x)$ is holomorphic in the spacetime coordinates $x^\mu$. To verify that it satisfies the field equation \eqref{eq:scalar_eq}, it is helpful to shift the $x$-dependence in \eqref{eq:scalar_transform} from the integration range into the integrand. This can be done by substituting $Z^a = P_\pm{}^a{}_b(x) W^b$, where $W^a$ is now integrated over an $x$-independent pair of Riemann spheres $P_\pm(y)$:
\begin{align}
 \begin{split}
   \varphi(x) ={}& \int_{P_-(y)}f_-\big(P_-(x) W\big) \wedge P_-{}^a{}_b(x) W_a dW^b \\
    &+ \int_{P_+(y)} f_+\big(P_+(x) W\big) \wedge P_+{}^a{}_b(x) W_a dW^b \ .
 \end{split} \label{eq:scalar_W}
\end{align}
This change of variables is regular, as long as $x$ is not null-separated from the antipode of $y$. The field equation \eqref{eq:scalar_eq} is now easy to verify, as detailed in Appendix \ref{app:field_eqs}. 

The main novelty of the transform \eqref{eq:scalar_transform} as compared to the Minkowski case is the presence of \emph{two} Riemann spheres, and with them two twistor functions. For the scalar field, however, the two separate integrals in \eqref{eq:scalar_transform} are redundant. It's sufficient to show this for the elementary functions \eqref{eq:f_elem}. Consider a function $f_-(Z)$ of the type \eqref{eq:f_elem}, normalized so that $A^a B_a = 1$. Then there exists a point $x'\in dS_{4,\bbC}$ such that $x'_\mu x'^\mu = 1$ and $2A^{[a} B^{b]} = P_+^{ab}(x')$. As shown in Appendix \ref{app:elementary}, the $P_-(x)$ piece of the transform \eqref{eq:scalar_transform} evaluates on the de Sitter hyperboloid $x_\mu x^\mu = 1$ as:
\begin{align}
 \varphi(x) = -\frac{4\pi i}{1 - x'_\mu x^\mu} = -\frac{8\pi i}{(x_\mu - x'_\mu)(x^\mu - x'^\mu)} \ .
\end{align}
It is now clear how the same field can be obtained from an integral over $P_+(x)$ rather than $P_-(x)$. One must simply choose a different function of the form \eqref{eq:f_elem}, such that $2A^{[a} B^{b]}$ equals $P_-^{ab}(x')$ rather than $P_+^{ab}(x')$. We conclude that just one of the $P_\pm(x)$ integrals is sufficient to obtain all the solutions \eqref{eq:scalar_transform}. This will not be the case for fields with spin, as we will see below.

Even though the presence of two Riemann spheres in \eqref{eq:scalar_transform} is redundant, it is still useful for expressing symmetries of the field $\varphi(x)$. The two functions $f_\pm(Z)$ may be related by three kinds of reflection symmetries, which induce the following properties on $\varphi(x)$: \begin{align}
 f_+(Z) = \pm f_-(Z) \quad &\longrightarrow \quad \varphi(x) = \pm \varphi(-x) \ ; \label{eq:minus_x} \\
 \bar f_+(Z) = e^{i\theta}f_-(\bar Z) \quad &\longrightarrow \quad \bar\varphi(x) = e^{i\theta}\varphi(\bar x) \ ; \label{eq:x_conj} \\
 \bar f_\pm(Z) = e^{i\theta}f_\pm(\bar Z) \quad &\longrightarrow \quad \bar\varphi(x) = e^{i\theta}\varphi(-\bar x) \ .
   \label{eq:minus_x_conj}
\end{align}
Here, $\theta$ is an arbitrary phase. The complex conjugations of forms and their arguments are understood as follows:
\begin{align}
 f(Z) = f_a(Z) d\bar Z^a \ ; \quad f(\bar Z) = f_a(\bar Z) d\Bar{\Bar Z}^a = -f_a(\bar Z) dZ^a \ ; \quad 
 \bar f(Z) = -\bar f_a(Z)\, dZ^a \ .
\end{align}
Eqs. \eqref{eq:minus_x}-\eqref{eq:minus_x_conj} can be verified directly from the transform \eqref{eq:scalar_transform}, using \eqref{eq:P_antipodal}-\eqref{eq:P_conjugate}. For the symmetries \eqref{eq:x_conj}-\eqref{eq:minus_x_conj} that involve complex conjugation, the proof requires an anti-holomorphic change of variables $Z^a\rightarrow \bar Z^a$. 

Two special cases of the symmetries \eqref{eq:minus_x}-\eqref{eq:minus_x_conj} should be noted. First, we see from \eqref{eq:x_conj} that for $\bar f_+(Z) = f_-(\bar Z)$, the field $\varphi(x)$ is real at real points $x$. Second, the symmetry \eqref{eq:minus_x} has a special status, since it doesn't involve complex conjugation (though it is consistent with the reality condition \eqref{eq:x_conj}). A field satisfying the holomorphic symmetry \eqref{eq:minus_x} can be written in terms of a single function $f_-(Z) \equiv f(Z)$ as:
\begin{align}
 \varphi(x) = \left(\int_{P_-(x)} \pm \int_{P_+(x)}\right) f(Z) Z_a dZ^a \ . \label{eq:scalar_Z2}
\end{align}
Such solutions can be viewed as holomorphic fields on $dS_{4,\bbC}/\bbZ_2$. The condition $\varphi(x) = \pm\varphi(-x)$ also follows from the discussion of charged fields in $dS_{4,\bbC}/\bbZ_2$ \cite{Parikh:2002py}, where charges at antipodal points must be opposite for consistency. Note that in \cite{Parikh:2002py}, the wrong symmetry $\varphi(x) = \pm\bar\varphi(-x)$ (for real $x$) was originally deduced instead of $\varphi(x) = \pm\varphi(-x)$. I thank Erik Verlinde for an email exchange on this point. 

\section{Free massless fields with spin} \label{sec:spin}

\subsection{Covariant derivatives of spinors in de Sitter space}

To discuss fields with spin, we need a convenient expression for covariant derivatives of spinors in the curved space $dS_4$. As with the d'Alembertian in section \ref{sec:scalar}, we will construct these from flat derivatives in $\bbR^{4,1}$. 

First, consider a left-handed spinor field $\varphi^\alpha(x)$ in de Sitter space. The Weyl index $\alpha$ can be upgraded into a Dirac index $a$, with zeros in the right-handed entries. As discussed in section \ref{sec:projectors:spinors}, this can also be viewed as a Dirac spinor index in $\bbR^{4,1}$, i.e. as a twistor index. Then on symmetry grounds, the following must be true:
\begin{align}
 \nabla_{\alpha\dot\beta}\,\varphi^\gamma = \gamma^\mu_{\alpha\dot\beta}\,\del_\mu\varphi^\gamma \ . \label{eq:nabla_left}
\end{align}
Here, the LHS is the covariant derivative in de Sitter space written with Weyl spinor indices. On the RHS, we have the flat derivative $\gamma^\mu_{ab}\del_\mu\varphi^c$, with the twistor indices projected onto the subspaces $P_\pm(x)$ to produce dotted/undotted Weyl indices. Similarly, for right-handed spinor fields $\varphi^{\dot\alpha}(x)$, we get:
\begin{align}
 \nabla_{\alpha\dot\beta}\,\varphi^{\dot\gamma} = \gamma^\mu_{\alpha\dot\beta}\,\del_\mu\varphi^{\dot\gamma} \ .
 \label{eq:nabla_right}
\end{align}

As with the scalar field before, we can extend $\varphi^a(x)$ away from the de Sitter hyperboloid $x_\mu x^\mu = 1$ by giving it a trivial radial dependence. However, this is not necessary: due to \eqref{eq:gamma_5}-\eqref{eq:gamma_4}, the projection $\gamma^\mu_{\alpha\dot\beta}$ of the Dirac indices in $\gamma^\mu_{ab}$ already selects the tangential components of the $\del_\mu$ derivative in \eqref{eq:nabla_left}-\eqref{eq:nabla_right}. 

The simple rules \eqref{eq:nabla_left}-\eqref{eq:nabla_right} are all we will need for the twistor transforms below. However, for completeness, let us also consider a Dirac field $\varphi^a(x)$ with both $\varphi^\alpha$ and $\varphi^{\dot\alpha}$ components. The covariant derivatives of these components are given by:
\begin{align}
 \begin{split}
   \nabla_{\alpha\dot\alpha}\,\varphi^\beta &= \gamma^\mu_{\alpha\dot\alpha}\,\del_\mu\varphi^\beta - i\delta_\alpha^\beta\,\varphi_{\dot\alpha} \ ; \\
   \nabla_{\alpha\dot\alpha}\,\varphi^{\dot\beta} &= \gamma^\mu_{\alpha\dot\alpha}\,\del_\mu\varphi^{\dot\beta} - i\delta_{\dot\alpha}^{\dot\beta}\,\varphi_{\alpha} \ . \label{eq:nabla_Dirac}
 \end{split}
\end{align}
The new terms on the RHS can again be deduced from symmetry, up to constant coefficients. The coefficient on e.g. the first line can be fixed by considering $\varphi^a(x) = P_+^{ab}(x)Z_b$ for a constant twistor $Z^a$. The covariant derivative on the LHS then vanishes, while the RHS is easy to evaluate. As a cross-check, one can verify that the derivatives \eqref{eq:nabla_Dirac} have the correct commutators for de Sitter space with unit radius:
\begin{align}
 \begin{split}
   &\nabla_{(\alpha}{}^{\dot\alpha}\nabla_{\beta)\dot\alpha} \varphi^\gamma = 2\delta_{(\alpha}^\gamma \varphi_{\beta)} \quad ; \quad 
   \nabla_{\alpha(\dot\alpha}\nabla^{\alpha}{}_{\dot\beta)} \varphi^\gamma = 0 \ ; \\
   &\nabla_{\alpha(\dot\alpha}\nabla^{\alpha}{}_{\dot\beta)} \varphi^{\dot\gamma} = -2\delta_{(\dot\alpha}^{\dot\gamma} \varphi_{\dot\beta)} \quad ; \quad
   \nabla_{(\alpha}{}^{\dot\alpha}\nabla_{\beta)\dot\alpha} \varphi^{\dot\gamma} = 0 \ .
 \end{split} \label{eq:commutators}
\end{align}

\subsection{The product-based twistor transform} \label{sec:spin:product}

Consider again the space of $\bar\del$-closed $(0,1)$-forms $f(Z)$ on twistor space defined up to $f\rightarrow f + \bar\del h$, this time with homogeneity $-2-n$ for positive integer $n$. For two such forms $f_\pm(Z)$, we define the transform:
\begin{align}
 \varphi_\pm^{a_1 a_2\dots a_n}(x) = \int_{P_\pm(x)} Z^{a_1}Z^{a_2}\dots Z^{a_n} f_\pm(Z)\wedge Z_b dZ^b \ , \label{eq:product_transform}
\end{align}
where the integrals are again over the Riemann spheres $P_\pm(x)$. The field $\varphi_\pm^{a_1\dots a_n}$ is clearly symmetric in all its indices. Since the $Z^a$ factors in \eqref{eq:product_transform} lie in $P_\pm(x)$, the Dirac indices on $\varphi_\pm^{a_1\dots a_n}$ are purely right-handed/left-handed. Thus, the only nonvanishing components of \eqref{eq:product_transform} are:
\begin{align}
 \begin{split}
   \varphi_-^{\alpha_1\alpha_2\dots\alpha_n}(x) &= \int_{P_-(x)} Z^{\alpha_1}Z^{\alpha_2}\dots Z^{\alpha_n} f(Z)\wedge Z_\beta dZ^\beta \ ; \\
   \varphi_+^{\dot\alpha_1\dot\alpha_2\dots\dot\alpha_n}(x) 
     &= \int_{P_+(x)} Z^{\dot\alpha_1}Z^{\dot\alpha_2}\dots Z^{\dot\alpha_n} f(Z)\wedge Z_{\dot\beta} dZ^{\dot\beta} \ .
 \end{split}
\end{align}
This implies that covariant derivatives of $\varphi_\pm^{\dots}(x)$ follow the Leibniz-rule extensions of eqs. \eqref{eq:nabla_left}-\eqref{eq:nabla_right}. One can then show that these fields satisfy the massless free field equations for helicity $\pm n/2$:
\begin{align}
 \nabla_{\alpha_1\dot\beta}\,\varphi_-^{\alpha_1\alpha_2\dots\alpha_n} = 0 \quad ; \quad 
 \nabla_{\beta\dot\alpha_1}\varphi_+^{\dot\alpha_1\dot\alpha_2\dots\dot\alpha_n} = 0 \ . \label{eq:product_eq}
\end{align}
The proof is similar to the scalar-field case, and is detailed in Appendix \ref{app:field_eqs}.

We see that the transform \eqref{eq:product_transform} produces free massless fields of \emph{both} left and right handedness from the same kind of twistor function. Unlike in the scalar-field case, the use of both Riemann spheres $P_\pm(x)$ is not redundant, since they produce fields with different handedness. The two functions $f_\pm(Z)$ can again be related by reflection symmetries, yielding the following relations for the fields $\varphi_\pm^{a_1\dots a_n}(x)$:
\begin{align}
 f_+(Z) &= \pm f_-(Z) &\longleftrightarrow&& \varphi_+^{a_1\dots a_n}(x) &= \pm \varphi_-^{a_1\dots a_n}(-x) \ ; 
   \label{eq:product_minus_x} \\
 \bar f_+(Z) &= e^{i\theta}f_-(\bar Z) &\longleftrightarrow&& 
   \bar\varphi_+^{a_1\dots a_n}(x) &= e^{i\theta}\varphi_-^{a_1\dots a_n}(\bar x) \ ; \label{eq:product_x_conj} \\
 \bar f_\pm(Z) &= e^{i\theta}f_\pm(\bar Z) &\longleftrightarrow&& 
   \bar\varphi_\pm^{a_1\dots a_n}(x) &= e^{i\theta}\varphi_\pm^{a_1\dots a_n}(-\bar x) \quad (\text{for }n\text{ even}) \ ,
   \label{eq:product_minus_x_conj} 
\end{align}
where the relations on $f_\pm(Z)$ are of course up to the freedom $f\rightarrow f + \bar\del h$. The restriction to even $n$ in \eqref{eq:product_minus_x_conj} is due to the anti-idempotence $\Bar{\Bar Z}^a = -Z^a$. We keep twistor indices on the fields in \eqref{eq:product_minus_x}-\eqref{eq:product_minus_x_conj}, since the symmetries relate spinors at \emph{different points} in de Sitter space. This makes them easier to express with twistor rather than Weyl-spinor indices, since the former are global, while the latter are local in $x$. From the Weyl-spinor point of view, eqs. \eqref{eq:product_minus_x}-\eqref{eq:product_minus_x_conj} make use of the isomorphisms between spinor spaces at $x$, $-x$ and $\bar x$, described in section \ref{sec:projectors:spinors}.

As in the scalar case, eq. \eqref{eq:product_x_conj} with $\theta=0$ and $x^\mu$ real implies that $\varphi_\pm^{\dots}(x)$ are the components of a real field. With Weyl-spinor indices, this condition reads: 
\begin{align}
 \bar\varphi_+^{\alpha_1\dots\alpha_n}(x) = \varphi_-^{\alpha_1\dots\alpha_n}(x) \quad ; \quad
 \bar\varphi_-^{\dot\alpha_1\dots\dot\alpha_n}(x) = (-1)^n\varphi_+^{\dot\alpha_1\dots\dot\alpha_n}(x) \ , \label{eq:product_real}
\end{align}
where the $(-1)^n$ is again due to the anti-idempotence $\Bar{\Bar Z}^a = -Z^a$.

Let us now discuss the holomorphic symmetry \eqref{eq:product_minus_x}. It equates the right-handed field $\varphi_+^{\dots}$ at $x$ with the left-handed field $\varphi_-^{\dots}$ at $-x$. In analogy with the scalar case, we conclude that such $\varphi_+^{\dots}(x)$ and $\varphi_-^{\dots}(x)$ can be combined into a spin-$n$ massless field on $dS_{4,\bbC}/\bbZ_2$. In Dirac-index notation, this field can be written as:
\begin{align}
 \varphi^{a_1\dots a_n}(x) = \left(\int_{P_-(x)} \pm \int_{P_+(x)}\right) Z^{a_1}\dots Z^{a_n} f(Z)\wedge Z_b dZ^b \ , \label{eq:product_Z2}
\end{align}
where $f(Z) \equiv f_-(Z)$. Since $dS_{4,\bbC}/\bbZ_2$ doesn't have a spacetime orientation, the field cannot have a definite handedness. Moreover, even in local orientable neighborhoods, the field cannot be purely left-handed or right-handed. Indeed, since $\varphi_+^{\dots}(x)$ and $\varphi_-^{\dots}(x)$ are holomorphic in $x$, neither of them can vanish in a neighborhood without vanishing everywhere. But under the symmetry \eqref{eq:product_minus_x}, if one of the two vanishes everywhere, then so must the other. We conclude that a non-vanishing holomorphic field on $dS_{4,\bbC}/\bbZ_2$ must have both left-handed and right-handed components in the neighborhood of every point. 

The reality condition \eqref{eq:product_real} and the holomorphic symmetry \eqref{eq:product_minus_x} are compatible when $n$ is even. Thus, on elliptical de Sitter space $dS_4/\bbZ_2$, only fields with integer spin may be real. 
 
\subsection{The derivative-based twistor transform} \label{sec:spin:deriv}

Finally, we turn to $\bar\del$-closed $(0,1)$-forms $f(Z)$ with homogeneity $-2+n$ for positive integer $n$. For two such forms $f_\pm(Z)$, we define the transform: 
\begin{align}
 \varphi^{(\pm)}_{a_1 a_2\dots a_n}(x) = \int_{P_\pm(x)} \frac{\del^n f_\pm(Z)}{\del Z^{a_1}\del Z^{a_2}\dots\del Z^{a_n}} \wedge Z_b dZ^b \ ,
 \label{eq:deriv_transform_raw}
\end{align}
where the integrals are again over the Riemann spheres $P_\pm(x)$, and the partial derivatives $\del/\del Z^a$ should not be confused with the holomorphic exterior derivative. The fields $\varphi^{(\pm)}_{a_1\dots a_n}$ are symmetric in all their indices. The partial derivatives in \eqref{eq:deriv_transform_raw} do not have a definite handedness: even though the integration variable $Z^a$ belongs to one of the $P_\pm(x)$ subspaces, the derivative $\del/\del Z^a$ does not. Nevertheless, the integral over each Riemann sphere picks out a single handedness component:
\begin{align}
 \begin{split}
   \varphi^{(-)}_{\dot\alpha_1\dot\alpha_2\dots\dot\alpha_n}(x) 
    &= \int_{P_-(x)} \frac{\del^n f_-(Z)}{\del Z^{\dot\alpha_1}\del Z^{\dot\alpha_2}\dots\del Z^{\dot\alpha_n}} \wedge Z_\beta dZ^\beta \ ; \\
   \varphi^{(+)}_{\alpha_1\alpha_2\dots\alpha_n}(x) 
    &= \int_{P_+(x)} \frac{\del^n f_+(Z)}{\del Z^{\alpha_1}\del Z^{\alpha_2}\dots\del Z^{\alpha_n}} \wedge Z_{\dot\beta} dZ^{\dot\beta} \ ,
 \end{split} \label{eq:deriv_transform} 
\end{align}
with all other components of $\varphi^{(\pm)}_{a_1\dots a_n}$ vanishing. Indeed, the integrand in e.g. $\varphi^{(-)}_{\alpha_1 a_2\dots a_n}$ can be rewritten as a total derivative, using the identity:
\begin{align}
 \ddsimple{F}{Z^\alpha}\wedge Z_\beta dZ^\beta = \ddsimple{F}{Z^\beta}\wedge Z_\alpha dZ^\beta - \ddsimple{F}{Z^\beta}\wedge Z^\beta dZ_\alpha 
  = -Z_\alpha dF + F\wedge dZ_\alpha = -d(F Z_\alpha) \ , \label{eq:F_alpha}
\end{align}
which holds for any holomorphic form $F(Z)$ on $P_-(x)$ with homogeneity $-1$. In the first equality in \eqref{eq:F_alpha}, we used the Fierz identity, exploiting the two-dimensionality of the subspace $P_-(x)$.   

As we demonstrate in Appendix \ref{app:field_eqs}, the fields \eqref{eq:deriv_transform} satisfy the massless free field equations for helicity $\pm n/2$:
\begin{align}
 \nabla_\beta{}^{\dot\alpha_1}\varphi^{(-)}_{\dot\alpha_1\dot\alpha_2\dots\dot\alpha_n} = 0 \quad ; \quad
 \nabla^{\alpha_1}{}_{\dot\beta}\,\varphi^{(+)}_{\alpha_1\alpha_2\dots\alpha_n} = 0 \ . \label{eq:deriv_eq}
\end{align}
The discussion of reality conditions and reflection symmetries is the same as in section \ref{sec:spin:product}. In particular, fields on $dS_{4,\bbC}/\bbZ_2$ can be constructed as:
\begin{align}
 \varphi_{a_1 a_2\dots a_n}(x) 
  = \left(\int_{P_-(x)} \pm \int_{P_+(x)}\right) \frac{\del^n f_\pm(Z)}{\del Z^{a_1}\del Z^{a_2}\dots\del Z^{a_n}} \wedge Z_b dZ^b \ .
 \label{eq:deriv_Z2}
\end{align}
For $n=2,4$, the fields \eqref{eq:deriv_transform} can be interpreted as the left-handed and right-handed components of a Maxwell field strength or a linearized Weyl curvature perturbation, respectively. For this interpretation to be consistent on $dS_{4,\bbC}/\bbZ_2$, one must choose the $+$ sign in \eqref{eq:deriv_Z2}.

\section{Discussion} \label{sec:discuss}

In this paper, we worked out the basics of twistor theory from a de Sitter-based perspective. Our main results are the twistor transforms \eqref{eq:product_transform},\eqref{eq:deriv_transform} that generate free massless fields on global de Sitter space $dS_4$, along with their counterparts \eqref{eq:product_Z2},\eqref{eq:deriv_Z2} for elliptical de Sitter space $dS_4/\bbZ_2$. These transforms arguably provide the most convenient method for constructing free massless solutions in these spacetimes. In Minkowski space, the situation is different. There, one can easily construct free solutions using momentum modes, making twistors truly useful only in the interacting theory. The same is true in the Poincare patch of de Sitter space, which is related to Minkowski space by a conformal transformation. Even for non-conformal fields, the spatial part of momentum modes in the Poincare patch remains trivial. On the other hand, in global (or elliptical) de Sitter space, the spatial part of momentum modes is replaced by spherical harmonics on $S_3$. This makes the twistorial method for constructing solutions competitive with the direct one.

More ambitiously, one could try for non-linear versions of the $dS_4/\bbZ_2$ transform \eqref{eq:deriv_Z2} for $n=2,4$, which would describe interacting Yang-Mills theory and gravity. One route is to look at non-perturbative classical solutions, as in the Penrose-Ward transform \cite{Ward:1977ta} and the non-linear graviton construction \cite{Penrose:1976js,Ward:1980am}. Another route is to construct a perturbation theory, such as the one utilized in the modern S-matrix calculations \cite{Adamo:2013cra}. A twistor description can be expected to shed light on field theory in $dS_4/\bbZ_2$, which, as motivated in \cite{Parikh:2002py}, is ultimately of interest for quantum gravity with positive cosmological constant. 

An important feature of our twistor transforms in $dS_4/\bbZ_2$ is that they come with no global notion of handedness. Thus, a non-linear version will not be restricted to self-dual fields, and in particular will include real solutions. More precisely, the hope is that this will be possible without squaring the twistor space, as one does in the ambitwistor approach \cite{Witten:1978xx,Isenberg:1978kk,Baston:1987av}. From the perturbative perspective, the lack of handedness implies that there is just one kind of external state, instead of two separate helicity signs. We note that non-linear versions of \eqref{eq:deriv_Z2} may turn out to be quite different from the standard constructions. For instance, as discussed in the Introduction, the infinity twistor $I_{ab}$ is a legitimate part of the conformal structure on $dS_4/\bbZ_2$. Thus, one may have a construction for Yang-Mills theory (which is classically conformal) that utilizes $I_{ab}$, in contrast with the standard wisdom. 

A major question concerning non-linear versions of \eqref{eq:deriv_Z2} would be the role of supersymmetry. In modern work on scattering amplitudes in Minkowski space, one evades the restriction to self-dual fields in twistor theory by invoking maximal supersymmetry, which puts both helicity signs in the same supermultiplet. On the other hand, in de Sitter space, one cannot have supersymmetry in the usual sense, due to the absence of a global timelike Killing vector. For $dS_4/\bbZ_2$, we can see two (mutually compatible) scenarios. First, as suggested in \cite{Parikh:2002py}, there may exist an adjusted notion of supersymmetry, once the lack of global time-orientation is correctly taken into account. Second, as implied above, it may turn out that supersymmetry is \emph{unnecessary} in $dS_4/\bbZ_2$, since there is no global distinction between helicity signs.

Finally, we note that our constructions can be carried over to anti-de Sitter space, at the price of some minus signs and factors of $i$. In fact, the complexified versions of de Sitter and AdS space are the same. However, this is only the case if one doesn't unwrap the periodic time coordinate in AdS. Global causality is then violated, more badly so than in $dS_4/\bbZ_2$, where closed timelike loops must pass through $\scri$. 

\section*{Acknowledgements}		

I am grateful to Abhay Ashtekar, Norbert Bodendorfer, Beatrice Bonga and Wolfgang Wieland for discussions, as well as to Lionel Mason and Erik Verlinde for email exchanges. This work is supported in part by the NSF grant PHY-1205388 and the Eberly Research Funds of Penn State.

\appendix
\section{Spacetime derivatives of the $P_\pm(x)$ projectors} \label{app:derivatives}

We list here some formulas for the derivatives $\del_\mu\equiv \del/\del x^\mu$ of the projector $P_-^{ab}(x)$, defined away from the de Sitter hyperboloid $x_\mu x^\mu = 1$ as in \eqref{eq:projectors_full}:
\begin{align}
 \del_\mu P_-^{ab} &= -\frac{i}{2\sqrt{x\cdot x}}\left(\delta_\mu^\nu - \frac{x_\mu x^\nu}{x\cdot x} \right) \gamma_\nu^{ab} \ ; \\
 \del_\mu\del^\mu P_-^{ab} &= \frac{2}{x\cdot x}\left(P_+^{ab} - P_-^{ab} \right) \ ; \\
 \del_\mu P_-^{ab} \del^\mu P_{-cd} &= \frac{1}{x\cdot x}\left(P_-{}^{[a}{}_c P_+{}^{b]}{}_d + P_+{}^{[a}{}_c P_-{}^{b]}{}_d \right) \ ; \\
 \gamma^\mu_{cd}\, \del_\mu P_-^{ab} &= \frac{2i}{\sqrt{x\cdot x}}\left(P_-{}^{[a}{}_c P_+{}^{b]}{}_d + P_+{}^{[a}{}_c P_-{}^{b]}{}_d \right) \ .
\end{align}
The derivatives of $P_+^{ab}(x)$ now follow from the relation \eqref{eq:P_antipodal}:
\begin{align}
 \del_\mu P_+^{ab} &= \frac{i}{2\sqrt{x\cdot x}}\left(\delta_\mu^\nu - \frac{x_\mu x^\nu}{x\cdot x} \right) \gamma_\nu^{ab} \ ; \\
 \del_\mu\del^\mu P_+^{ab} &= \frac{2}{x\cdot x}\left(P_-^{ab} - P_+^{ab} \right) \ ; \\
 \del_\mu P_+^{ab} \del^\mu P_{+cd} &= \frac{1}{x\cdot x}\left(P_-{}^{[a}{}_c P_+{}^{b]}{}_d + P_+{}^{[a}{}_c P_-{}^{b]}{}_d \right) \ ; \\
 \gamma^\mu_{cd}\, \del_\mu P_+^{ab} &= -\frac{2i}{\sqrt{x\cdot x}}\left(P_-{}^{[a}{}_c P_+{}^{b]}{}_d + P_+{}^{[a}{}_c P_-{}^{b]}{}_d \right) \ .
\end{align}

The above formulas are useful for taking derivatives under the integral sign in the twistor transform \eqref{eq:scalar_W} and its counterparts with nonzero spin. More specifically, we are interested there in combinations of the form $Z_\pm^a(x) = P_\pm{}^a{}_b(x) Z^b$, with constant $Z^a$. For $Z^a_-(x)$, we get:
\begin{align}
 \del_\mu Z_-^a &= \frac{i}{2\sqrt{x\cdot x}}\left(\delta_\mu^\nu - \frac{x_\mu x^\nu}{x\cdot x} \right) \gamma_\nu^{ab} Z_b \ ; \\
 \del_\mu\del^\mu Z_-^a &= \frac{2}{x\cdot x}\left(Z_+^a - Z_-^a \right) \ ; \\
 \del_\mu Z_-^a \del^\mu Z_-^b &= \frac{1}{x\cdot x}\, Z_-^{(a} Z_+^{b)} \ ; \\
 \del_\mu Z_-^a \del^\mu\tilde Z_-^b &= \frac{1}{2(x\cdot x)}\left(\tilde Z_-^a Z_+^b + \tilde Z_+^a Z_-^b
   + P_-^{ab}\tilde Z_{+c}Z_+^c + P_+^{ab}\tilde Z_{-c}Z_-^c \right) \ ; \\ 
 \del_\mu Z_{-a} \del^\mu\tilde Z_-^a &= \frac{1}{x\cdot x}\left(Z_{-a} \tilde Z_-^a + Z_{+a} \tilde Z_+^a \right) 
   = \frac{1}{x\cdot x}\,Z_a\tilde Z^a \ ; \\
 \gamma^{\mu bc}\del_\mu Z_-^a &= -\frac{2i}{\sqrt{x\cdot x}}\left(P_-^{a[b}Z_+^{c]} + P_+^{a[b}Z_-^{c]} \right) \ , 
\end{align}   
where $\tilde Z^a$ is some other constant twistor, and $\tilde Z^a_-(x) \equiv P_-{}^a{}_b(x)\tilde Z^b$. Similarly, for $Z_+^a(x)$ and $\tilde Z_+^a(x)$, we get:
\begin{align}
 \del_\mu Z_+^a &= -\frac{i}{2\sqrt{x\cdot x}}\left(\delta_\mu^\nu - \frac{x_\mu x^\nu}{x\cdot x} \right) \gamma_\nu^{ab} Z_b \ ; \\
 \del_\mu\del^\mu Z_+^a &= \frac{2}{x\cdot x}\left(Z_-^a - Z_+^a \right) \ ; \\
 \del_\mu Z_+^a \del^\mu Z_+^b &= \frac{1}{x\cdot x}\, Z_-^{(a} Z_+^{b)} \ ; \\
 \del_\mu Z_+^a \del^\mu\tilde Z_+^b &= \frac{1}{2(x\cdot x)}\left(\tilde Z_-^a Z_+^b + \tilde Z_+^a Z_-^b
   + P_-^{ab}\tilde Z_{+c}Z_+^c + P_+^{ab}\tilde Z_{-c}Z_-^c \right) \ ; \\ 
 \del_\mu Z_{+a} \del^\mu\tilde Z_+^a &= \frac{1}{x\cdot x}\left(Z_{-a} \tilde Z_-^a + Z_{+a} \tilde Z_+^a \right) 
  = \frac{1}{x\cdot x}\,Z_a\tilde Z^a \ ; \\
 \gamma^{\mu bc}\del_\mu Z_+^a &= \frac{2i}{\sqrt{x\cdot x}}\left(P_-^{a[b}Z_+^{c]} + P_+^{a[b}Z_-^{c]} \right) \ . 
\end{align}   

\section{Deriving the field equations} \label{app:field_eqs}

Here, we verify that the twistor transforms \eqref{eq:scalar_transform}, \eqref{eq:product_transform} and \eqref{eq:deriv_transform} satisfy the free field equations \eqref{eq:scalar_eq}, \eqref{eq:product_eq} and \eqref{eq:deriv_eq}, respectively.

\subsection{Scalar field}

Consider the $P_-(x)$ piece of the scalar-field transform \eqref{eq:scalar_transform}:
\begin{align}
 \varphi(x) = \int_{P_-(x)} f(Z)\wedge Z_a dZ^a \ .
\end{align}
Let us show that it satisfies the field equation \eqref{eq:scalar_eq}. The proof for the $P_+(x)$ piece is similar. First, rewrite the integral as in \eqref{eq:scalar_W}, moving the $x$-dependence into the integrand:
\begin{align}
 \varphi(x) ={}& \int_{P_-(y)}f\big(P_-(x) Z\big) \wedge P_-{}^a{}_b(x) Z_a\, dZ^b 
   = \int f(Z_-) \wedge Z_{-a} dZ_-^a \ . 
 \label{eq:scalar_transform_app}
\end{align}
$Z^a$ is now integrated over a fixed Riemann sphere $P_-(y)$, and we denote $Z_\pm^a \equiv P_\pm{}^a{}_b(x) Z^b$. Using the derivative formulas from Appendix \ref{app:derivatives} with the identity \eqref{eq:gamma_gamma} and substituting $x_\mu x^\mu = 1$ at the end of the calculation, we find that the integrand satisfies:
\begin{align}
 \begin{split}
  &\del_\mu \del^\mu\left(f(Z_-)\wedge Z_{-a}dZ_-^a \right) = \\ 
   &\ = \left(\frac{\del^2 f}{\del Z_-^b\del Z_-^c}Z_-^b Z_+^c + 3\ddsimple{f}{Z_-^b}Z_+^b - 2\ddsimple{f}{Z_-^b}Z_-^b - 2f \right)\wedge Z_{-a}dZ_-^a \\ 
   &\ \quad + \left(\ddsimple{f}{Z_-^b}Z_-^b + 2f \right)\wedge Z_{+a}dZ_+^a \ . \label{eq:box_raw}
 \end{split}
\end{align}
We now use the homogeneity relations:
\begin{align}
 \ddsimple{f}{Z_-^a}Z_-^a = -2f \quad ; \quad \frac{\del^2 f}{\del Z_-^a\del Z_-^b}Z_-^b = -3\ddsimple{f}{Z_-^a} \ ,
\end{align}
which bring \eqref{eq:box_raw} to the form:
\begin{align}
 \del_\mu \del^\mu\left(f(Z_-)\wedge Z_{-a}dZ_-^a \right) = 2f(Z_-)\wedge Z_{-a}dZ_-^a \ .
\end{align}
Substituting into the integral \eqref{eq:scalar_transform_app}, we obtain the field equation:
\begin{align}
 \Box\varphi(x) = \del_\mu \del^\mu \varphi(x) = 2\varphi(x) \ .
\end{align}
 
\subsection{Spinor field from the product-based transform}

We now turn to the $P_-(x)$ piece of the transform \eqref{eq:product_transform}:
\begin{align}
 \varphi_-^{a_1 a_2\dots a_n}(x) = \int_{P_-(x)} Z^{a_1}Z^{a_2}\dots Z^{a_n} f(Z)\wedge Z_b dZ^b \ .
\end{align}
Let us show that it satisfies the field equation \eqref{eq:product_eq}. Again, the proof for the $P_+(x)$ piece is similar. Rewriting the integral as in \eqref{eq:scalar_transform_app} to move the $x$-dependence into the integrand, we get:
\begin{align}
 \varphi_-^{a_1 a_2\dots a_n}(x) = \int f(Z_-)Z_-^{a_1} Z_-^{a_2}\dots Z_-^{a_n} \wedge Z_{-b} dZ_-^b \ , 
 \label{eq:product_transform_app}
\end{align}
where $Z_\pm^a \equiv P_\pm{}^a{}_b(x) Z^b$, and $Z^a$ is integrated over a fixed Riemann sphere. Using the formulas from Appendix \ref{app:derivatives} with the identity \eqref{eq:gamma_gamma} and substituting $x_\mu x^\mu = 1$ at the end, we find that the integrand satisfies:
\begin{align}
 \begin{split}
  &P_-{}^c{}_{a_1}(x)\,\gamma^\mu_{cd}\,\del_\mu \left(f(Z_-)Z_-^{a_1} Z_-^{a_2}\dots Z_-^{a_n}\wedge Z_{-b}dZ_-^b \right) = \\ 
  &\ = -i\left(\ddsimple{f}{Z_-^c}Z_-^c + (n+2)f \right)Z_{+d} Z_-^{a_2}\dots Z_-^{a_n}\wedge Z_{-b}dZ_-^b \ . \label{eq:product_raw} 
 \end{split}
\end{align}
Using the homogeneity relation:
\begin{align}
 \ddsimple{f}{Z_-^a}Z_-^a = -(n+2)f
\end{align}
and translating into Weyl-spinor indices, eq. \eqref{eq:product_raw} becomes:
\begin{align}
 \gamma^\mu_{\alpha_1\dot\beta}\,\del_\mu \left(f(Z_-)Z_-^{\alpha_1} Z_-^{\alpha_2}\dots Z_-^{\alpha_n}\wedge Z_{-\gamma}dZ_-^\gamma \right) = 0 \ . 
\end{align}
Substituting into the integral \eqref{eq:product_transform_app} and using the recipe \eqref{eq:nabla_left} for covariant derivatives, we obtain the field equation:
\begin{align}
 \nabla_{\alpha_1\dot\beta}\,\varphi_-^{\alpha_1\alpha_2\dots\alpha_n} 
  = \gamma^\mu_{\alpha_1\dot\beta}\,\del_\mu\varphi_-^{\alpha_1\alpha_2\dots\alpha_n} = 0 \ .
\end{align}

\subsection{Spinor field from the derivative-based transform}

Finally, consider the $P_-(x)$ piece of the transform \eqref{eq:deriv_transform}:
\begin{align}
 \varphi^{(-)}_{a_1 a_2\dots a_n}(x) = \int_{P_-(x)} \frac{\del^n f(Z)}{\del Z^{a_1}\del Z^{a_2}\dots\del Z^{a_n}}\wedge Z_b dZ^b \ .
\end{align}
Let us show that it satisfies the field equation \eqref{eq:deriv_eq}. Rewriting the integral as in \eqref{eq:scalar_transform_app} to move the $x$-dependence into the integrand, we get:
\begin{align}
 \varphi^{(-)}_{a_1 a_2\dots a_n}(x) = \int \frac{\del^n f(Z_-)}{\del Z_-^{a_1}\del Z_-^{a_2}\dots\del Z_-^{a_n}}\wedge Z_{-b} dZ_-^b \ , 
 \label{eq:deriv_transform_app}
\end{align}
where $Z_\pm^a \equiv P_\pm{}^a{}_b(x) Z^b$, and $Z^a$ is integrated over a fixed Riemann sphere. Using the formulas from Appendix \ref{app:derivatives} with the identity \eqref{eq:gamma_gamma} and substituting $x_\mu x^\mu = 1$ at the end, we find that the integrand satisfies:
\begin{align}
 \begin{split}
  &P_+{}^{a_1}{}_d(x)\,\gamma_\mu^{cd}\,\del^\mu \left(\frac{\del^n f(Z_-)}{\del Z_-^{a_1}\del Z_-^{a_2}\dots\del Z_-^{a_n}} 
   \wedge Z_{-b}dZ_-^b \right) = \\ 
  &= i\left(P_-^{ce} \frac{\del^{n+1} f(Z_-)}{\del Z_-^e\del Z_-^d\del Z_-^{a_2}\dots\del Z_-^{a_n}} \wedge Z_+^d Z_{-b}dZ_-^b 
  + \frac{\del^n f(Z_-)}{\del Z_-^d\del Z_-^{a_2}\dots\del Z_-^{a_n}}\wedge \left(Z_+^d dZ_-^c - Z_-^c dZ_+^d \right) \right) \ .
 \label{eq:deriv_raw} 
 \end{split}
\end{align}
On the fixed Riemann sphere where the integration variable $Z^a$ lives, the functions $Z^a_\pm$ are linearly related as $Z_+^a = L^a{}_b Z_-^b$, where the matrix $L^a{}_b$ depends on the fixed Riemann sphere and on $x^\mu$. Eq. \eqref{eq:deriv_raw} then becomes:
\begin{align}
 \begin{split}
  &P_+{}^{a_1}{}_d(x)\,\gamma_\mu^{cd}\,\del^\mu \left(\frac{\del^n f(Z_-)}{\del Z_-^{a_1}\del Z_-^{a_2}\dots\del Z_-^{a_n}} 
   \wedge Z_{-b}dZ_-^b \right) = \\ 
  &= iL^d{}_m\left(P_-^{ce}(x)\, \frac{\del^{n+1} f(Z_-)}{\del Z_-^e\del Z_-^d\del Z_-^{a_2}\dots\del Z_-^{a_n}} \wedge Z_-^m Z_{-b}dZ_-^b 
  + \frac{\del^n f(Z_-)}{\del Z_-^d\del Z_-^{a_2}\dots\del Z_-^{a_n}}\wedge \left(Z_-^m dZ_-^c - Z_-^c dZ_-^m \right) \right) \ .
 \label{eq:deriv_raw_2} 
 \end{split}
\end{align}
We now invoke the identity \eqref{eq:F_alpha} in the form:
\begin{align}
 P_-^{ce}(x)\frac{\del}{\del Z_-^e}\left(\frac{\del^n f(Z_-)}{\del Z_-^d\del Z_-^{a_2}\dots\del Z_-^{a_n}}\, Z_-^m \right)\wedge Z_{-b}dZ_-^b 
  = -d\left(\frac{\del^n f(Z_-)}{\del Z_-^d\del Z_-^{a_2}\dots\del Z_-^{a_n}}\, Z_-^m Z_-^c \right) \ .
\end{align}
This allows us to rewrite \eqref{eq:deriv_raw_2} as:
\begin{align}
 \begin{split}
  &P_+{}^{a_1}{}_d(x)\,\gamma_\mu^{cd}\,\del^\mu \left(\frac{\del^n f(Z_-)}{\del Z_-^{a_1}\del Z_-^{a_2}\dots\del Z_-^{a_n}} 
   \wedge Z_{-b}dZ_-^b \right) = \\ 
  &\qquad = iL^d{}_m \frac{\del^n f(Z_-)}{\del Z_-^d\del Z_-^{a_2}\dots\del Z_-^{a_n}}\wedge 
   \left(-P_-^{cm}(x) Z_{-b}dZ_-^b + Z_-^m dZ_-^c - Z_-^c dZ_-^m \right) \\
  &\qquad \quad -iL^d{}_m d\left(\frac{\del^n f(Z_-)}{\del Z_-^d\del Z_-^{a_2}\dots\del Z_-^{a_n}}\, Z_-^m Z_-^c \right) \\
  &\qquad = 0 - iL^d{}_m d\left(\frac{\del^n f(Z_-)}{\del Z_-^d\del Z_-^{a_2}\dots\del Z_-^{a_n}}\, Z_-^m Z_-^c \right) \ .
 \label{eq:deriv_raw_3} 
 \end{split} 
\end{align}
where the last step follows from the Fierz identity in the 2d subspace $P_-(x)$. Integrating eq. \eqref{eq:deriv_raw_3} over the Riemann sphere, we get:
\begin{align}
 P_+{}^{a_1}{}_d(x)\,\gamma_\mu^{cd}\,\del^\mu \varphi^{(-)}_{a_1 a_2\dots a_n}(x) = 0 \ .
\end{align}
Translating into Weyl-spinor indices, projecting $a_2\dots a_n$ onto the relevant subspace and using the recipe \eqref{eq:nabla_right} for covariant derivatives, we finally obtain the field equation:
\begin{align}
 \nabla^{\beta\dot\alpha_1}\,\varphi^{(-)}_{\dot\alpha_1\dot\alpha_2\dots\dot\alpha_n} 
  = \gamma_\mu^{\beta\dot\alpha_1}\,\del^\mu\varphi^{(-)}_{\dot\alpha_1\dot\alpha_2\dots\dot\alpha_n} = 0 \ .
\end{align}

\section{The scalar-field transform of an elementary twistor function} \label{app:elementary}

Here, we calculate the scalar-field twistor transform \eqref{eq:scalar_transform} for a function $f_-(Z)$ of the form \eqref{eq:f_elem}, where $2A^{[a}B^{b]} = P_+^{ab}(x')$ for some point $x'^\mu$ with $x'_\mu x'^\mu = 1$. The integration variable $Z^a$ can be decomposed in terms of two complex numbers $w \equiv A_a Z^a$ and $z \equiv B_a Z^a$:
\begin{align}
 Z^a = \frac{1}{P_-^{cd}(x)A_c B_d}\left(w P_-^{ab}(x)B_b - z P_-^{ab}(x)A_b \right) \ .
\end{align}
The pair $(w,z)$ act as homogeneous coordinates on a Riemann sphere $\bbC\bbP^1$. The coordinate transformation from $Z^a$ on $P_-(x)$ to $(w,z)$ on $\bbC\bbP^1$ is regular, as long as $x$ and $x'$ are not null-separated. The measure $Z_a dZ^a$ becomes:
\begin{align}
 Z_a dZ^a = \frac{zdw - wdz}{P_-^{ab}(x)A_a B_b} = \frac{2(zdw - wdz)}{P_-^{ab}(x)P_{+ab}(x')} = \frac{2(zdw - wdz)}{1 - x'_\mu x^\mu} \ .
\end{align}
The transform \eqref{eq:scalar_transform} then reads: 
\begin{align}
 \begin{split}
  \varphi(x) &= \int_{P_-(x)} f_-(Z)\wedge Z_a dZ^a 
     = \int_{P_-(x)} \frac{1}{A_a Z^a}\, \bar\del\, \frac{1}{B_b Z^b} \wedge Z_c dZ^c \\
   &= \frac{2}{1 - x'_\mu x^\mu} \int_{\bbC\bbP^1} \bar\del\,\frac{1}{z}\wedge \left(z\frac{dw}{w} - dz \right) 
   = \frac{2}{1 - x'_\mu x^\mu}(0 - 2\pi i) = -\frac{4\pi i}{1 - x'_\mu x^\mu}  \ .
 \end{split}  
\end{align}
Using $x_\mu x^\mu = x'_\mu x'^\mu = 1$, the result can be rewritten as:
\begin{align}
 \varphi(x) = -\frac{8\pi i}{(x_\mu - x'_\mu)(x^\mu - x'^\mu)} \ .
\end{align}
In this form, it's clear that the field is singular on the lightcone of $x'$.

\end{document}